\documentclass[]{aastex631}

\usepackage{amssymb}
\usepackage{graphicx}
\usepackage{amsmath}
\usepackage{aas_macros}

\begin{document}

\title{Locating the missing baryons in the warm-hot intergalactic medium with fast radio bursts and the Sunyaev-Zel'dovich effect}

\correspondingauthor{F.Y. Wang}
\email{fayinwang@nju.edu.cn}

\author[0000-0001-7176-8170]{Dao-Hong Zhai}
\affiliation{School of Astronomy and Space Science, Nanjing University, Nanjing 210093, China}

\author[0000-0003-4157-7714]{F.Y. Wang}
\affiliation{School of Astronomy and Space Science, Nanjing University, Nanjing 210093, China}
\affiliation{Key Laboratory of Modern Astronomy and Astrophysics (Nanjing University), Ministry of Education, Nanjing 210093, China}

\author{Zi-Gao Dai}
\affiliation{Department of Astronomy, School of Physical Sciences, University of Science and Technology of China, Hefei 230026, China}

\author{Renyue Cen}
\affiliation{Center for Cosmology and Computational Astrophysics, Zhejiang University, Hangzhou 310027, China}

\begin{abstract}

Traditional astronomical censuses in the late-time Universe can only account for a fraction of the baryonic matter budget. Hydrodynamical simulations predict that the missing baryons reside in the vast filamentary structures of the cosmic web as a highly diffuse, warm-hot intergalactic medium (WHIM). Observing the WHIM directly has remained a long-standing challenge due to its typical temperature. In this study, we measure the angular cross-power spectrum between the dispersion measures (DMs) of fast radio bursts (FRBs) from the second CHIME/FRB catalog and the thermal Sunyaev-Zel'dovich (tSZ) Compton-$y$ map from the Planck satellite. After masking cataloged galaxy clusters, we find a positive correlation between FRB DMs and the tSZ map at $>98\%$ confidence ($p_{\rm null}=0.0169$). With a temperature normalization of $2.4\times10^6\ {\rm K}$, we infer that a fraction $f_{\rm WHIM}=0.375^{+0.165}_{-0.161}$ of cosmic baryons resides in the hotter WHIM. These measurements constrain a substantial reservoir of hotter diffuse gas in the cosmic web and statistically close the cosmic baryon budget in the local Universe under the adopted density and pressure model.

\end{abstract}

\keywords{Fast radio bursts, Sunyaev-Zel'dovich effect, warm-hot intergalactic medium, missing baryons}

\section{Introduction}
\label{sec:introduction}
Precision measurements of the cosmic microwave background (CMB) and Big Bang nucleosynthesis have firmly established the total baryonic matter content of the Universe \citep{Aghanim2020, Cooke2018}. 
At moderate redshifts $z=2-4$, Lyman-alpha forest observations can account for the vast majority of the baryons inferred by CMB observations \citep{Weinberg1997, Rauch1997, Haehnelt2001, Hui2002, Tytler2004}.
However, traditional astronomical censuses in the late-time Universe reveal a long-standing puzzle known as the ``missing baryon problem'', i.e., the summation of all observed ordinary matter within stars, cold interstellar gas, and dense intra-halo environments accounts for merely a fraction of this expected cosmological budget \citep{Fukugita1998, Bregman2007, Shull2012, Macquart2020}. Hydrodynamical simulations predict that these baryons are not genuinely missing, but rather reside in the extended filamentary structures of the cosmic web as a highly diffuse and fully ionized plasma known as the warm-hot intergalactic medium (WHIM) \citep{Cen1999, Dave2001, Graaff2019, Tuominen2021}. Shock-heated to temperatures between $10^5 \ {\rm K}$ and $10^7 \ {\rm K}$, the WHIM is extremely difficult to observe directly. 
While a significant portion of the low-temperature ($T\le 3\times 10^5$K) WHIM has been detected in absorption in the ultraviolet \citep{Danforth2005}, the hotter portion of the WHIM is generally too faint to be easily detected through X-ray emission
\citep{Fang2002,Bregman2007,TepperGarcia2011}. Candidate detections have also been reported through X-ray absorption, although the interpretation can depend sensitively on the environment and redshift of the background source \citep{Nicastro2018,Johnson2019,DorigoJones2022}. New cosmological probes are urgently required to trace especially the hotter part of the WHIM to provide a wholistic picture of the missing baryons.

Fast radio bursts (FRBs) are millisecond-duration, highly energetic radio transients of extragalactic origin \citep{Lorimer2007,Petroff2022,Zhang2023,Wang2026}. The defining observational feature of an FRB is its dispersion measure (DM), which quantifies the frequency-dependent arrival time delay of the radio pulse. Physically, the DM represents the integrated column density of free electrons along the entire line of sight from the source to the observer. For a cosmological FRB located at redshift $z$, the total observed DM can be decomposed into several components:
\begin{equation}
    {\rm DM_{obs}} = {\rm DM_{MW}} + {\rm DM_{halo}} + {\rm DM_{IGM}} + \frac{{\rm DM_{host}}}{1+z},
\end{equation}
where ${\rm DM_{MW}}$ and ${\rm DM_{halo}}$ are the contributions from the Milky Way's interstellar medium (ISM) and its circumgalactic halo, respectively. The term ${\rm DM_{IGM}}$ arises from the diffuse intergalactic medium (IGM), and ${\rm DM_{host}}$ originates from the FRB host galaxy and its immediate local environment. Because the dominant contribution for distant FRBs typically comes from the IGM, these transients serve as unique and powerful cosmological probes \citep{Bhandari2022,Wu2024}. Unlike traditional absorption-line spectroscopy, which is highly sensitive to the ionization state and temperature of the gas, the FRB DM acts as an unbiased counter of all free electrons along the propagation path \citep{Munoz2018}. Recent studies have utilized this property of FRBs to measure the Hubble constant \citep{Wu2022,James2022,Gao2025} and the fraction of cosmic baryons residing in the IGM \citep{McQuinn2014,Macquart2020,Yang2022,Connor2025}. The statistical robustness of these cosmological applications relies heavily on the available sample size. In this regard, the Canadian Hydrogen Intensity Mapping Experiment (CHIME) serves as a revolutionary instrument detecting FRBs at an unprecedented rate \citep{Amiri2018,Amiri2021}. The recent release of the second CHIME/FRB catalog \citep{FRBCollaboration2026} (hereafter Catalog 2) comprises thousands of well-characterized events, establishing large samples of FRBs for various statistical studies.

While DMs of FRBs trace the total column density of electrons regardless of their thermodynamic state, the thermal Sunyaev-Zel'dovich (tSZ) effect provides a highly complementary probe sensitive to the thermal energy of the gas. The tSZ effect arises from the inverse-Compton scattering of CMB photons by high-energy ionized gas in the large-scale structure \citep{Sunyaev1972}. The magnitude of this signal is characterized by the dimensionless Compton-$y$ parameter, which is defined as the line-of-sight integral of the electron pressure:
\begin{equation}
    y = \int \frac{k_{\rm B} T_e}{m_e c^2} \sigma_{\rm T} n_e {\rm d}l = \int \frac{\sigma_{\rm T}}{m_e c^2} P_e {\rm d}l,
\end{equation}
where $k_{\rm B}$ is the Boltzmann constant, $T_e$ is the electron temperature, $m_e$ is the electron mass, $\sigma_{\rm T}$ is the Thomson scattering cross-section, $n_e$ is the electron number density, and $P_e$ is the electron pressure. Because the Compton-$y$ parameter is directly proportional to $P_e$, the tSZ effect has been extensively utilized to map the intracluster medium (ICM) and to catalog thousands of massive galaxy clusters \citep{Ade2016a}. Besides these virialized halos, the tSZ effect also offers an effective pathway to detect the fainter, extended diffuse gas in the cosmic web \citep{Komatsu2002,Graaff2019,Tanimura2020,Ibitoye2024,Li2025}. Notably, a $\sim 3\sigma$ tSZ signal was detected from the intergalactic filaments by stacking approximately one million pairs of luminous red galaxies \citep{Graaff2019}, providing clues for the existence of the WHIM at temperatures of $10^5 - 10^7 \ {\rm K}$.

Although FRBs are high-precision cosmological probes, their DMs measure the total electron column density regardless of the gas temperature, rendering them incapable of distinguishing between cold, warm, and hot gas phases. Another major challenge in traditional DM-based cosmological inferences is the degeneracy between ${\rm DM_{IGM}}$ and ${\rm DM_{host}}$, which requires additional assumptions on the free-electron distribution within host galaxies \citep{Macquart2020,Zhang2020}. Meanwhile, sparse FRB sampling and source-to-source DM scatter limit measurements of the DM auto-power spectrum \citep{Sharma2025}. For the sample and angular scales used here, our measured auto-power spectrum is dominated by noise. A recent study by \citet{Wang2026} reported a DM auto-correlation signal using Catalog 2. While the tSZ effect is highly sensitive to the gas temperature, its auto-power spectrum is predominantly driven by the ultra-hot ICM within massive galaxy clusters and suffers from severe systematic contamination from the Cosmic Infrared Background (CIB) \citep{Ibitoye2024}.

Because FRB DMs and the tSZ Compton-$y$ parameter trace electron density and pressure, their cross-correlation is especially sensitive to the hotter WHIM and probes its thermal state \citep{Munoz2018}. Uncorrelated noise does not contribute to the mean cross-power spectrum, but host environments and residual foregrounds may still produce correlated contributions. We examine independent host-DM scatter in Section~\ref{sec:host_tests} and compare different foreground treatments in Sections~\ref{sec:cib_tests} and \ref{sec:galactic_tests}. These tests assess the sensitivity of the inferred hotter WHIM fraction to host scatter and foreground treatment.

Recent studies have successfully cross-correlated both FRBs and the tSZ effect with various distinct large-scale structure tracers, such as galaxy catalogs, weak lensing shear, and the integrated Sachs-Wolfe effect \citep{Ibitoye2024, Wang2025, Leung2025, Sharma2025, Wayland2026}. The potential of directly cross-correlating FRBs with tSZ maps was first theoretically envisioned by \citet{Munoz2018}, who proposed that a tSZ map combined with high-resolution angular localization of FRBs could extract the weak WHIM signal through cross-correlation. Using localized FRBs with measured redshifts, \citet{Takahashi2025} measured a real-space DM-tSZ correlation from DM residuals and annular averages of the $y$ maps around the FRB positions. Notably, a concurrent study by \citet{Sharma2026} also reported a positive angular correlation between CHIME FRB DMs and various large-scale structure tracers, including the Planck tSZ map. Their tSZ measurement, which includes contributions from massive galaxy clusters, primarily constrains the strength of baryonic feedback. Extending these studies to underdense regions, \citet{Sharma2026void} combined a void DM deficit with existing tSZ stacks to constrain the mean void-gas temperature. In this work, we measure the angular cross-power spectrum between Catalog 2 FRB DMs and the cluster-masked tSZ map to target the faint, extended gas in the hotter WHIM. By masking cataloged galaxy clusters, we detect this cross-correlation at $>98\%$ confidence ($p_{\rm null}=0.0169$). Under the adopted temperature and pressure model, roughly $37.5\%$ of cosmic baryons reside in this hotter component.

\begin{figure}[!h]
\centering
\includegraphics[width=0.7\textwidth]{preview.png}
\caption{\textbf{Sky distribution of the FRB sample and the masked tSZ map.} \textbf{(a)} The spatial distribution of FRBs and galaxy clusters. Blue and green dots represent clusters from the PSZ2 and MCXC-II catalogs, respectively. Crosses indicate FRBs located within a $3\theta_{500}$ exclusion radius of any cluster and these are discarded. Orange dots represent the final FRB sample retained for the cross-correlation measurement. \textbf{(b)} The masked Compton-$y$ parameter map. Circular regions within a $3\theta_{500}$ radius around all cataloged clusters are excluded to isolate the hotter diffuse WHIM signal. The exclusion mask is smoothed using a cosine apodization to prevent ringing artifacts. The Galactic plane and bright point sources are also masked out.}
\label{fig:FRB_SZ_map}
\end{figure}

\section{Methods}
\label{sec:methods}

\subsection{Data selection and masking strategies}
\subsubsection{FRB and tSZ data}
We start from a sample of 2656 extragalactic FRBs before applying the cluster exclusions and the final $y$-map mask. We utilize the all-sky Compton-$y$ parameter map \citep{Chandran2023}, derived from the Planck PR4 frequency maps \citep{Akrami2020}. This map is generated via the Needlet Internal Linear Combination (NILC) method specifically tailored for tSZ signal extraction \citep{Delabrouille2009, Remazeilles2011,Remazeilles2013}, which minimizes the total error variance in both pixel and needlet domains. Compared to the PR2 $y$-map released by the Planck Collaboration \citep{Aghanim2016}, the PR4-based map exhibits significantly reduced instrumental noise and large-scale systematics, along with lower thermal dust residuals in the Galactic plane. The map is provided in the HEALPix pixelization scheme \citep{Gorski2005} at a resolution of $N_{\rm side} = 2048$, with an effective beam FWHM of 10 arcmin. We employ the three apodized masks provided in the work of \citet{Chandran2023}: the standard NILC processing mask (NILC-MASK, $f_{\rm sky} \approx 98\%$), the Galactic mask (GAL-MASK, $f_{\rm sky} \approx 60\%$, originally defined in \citet{Aghanim2016}), and the point-source mask (PS-MASK), which excises bright radio and infrared sources. We adopt the product of these three masks as our base tSZ mask, yielding a final sky fraction of $f_{\rm sky} \approx 56\%$.

We draw our FRB sample from CHIME/FRB Catalog 2 \citep{FRBCollaboration2026}, which reports 4539 FRBs from 3641 unique sources detected between 2018 and 2023. This release provides a substantial increase in sample size and improved calibration compared to the first catalog. Several data cleaning procedures are applied to ensure a high-quality dataset for cross-correlation analysis. We first remove all events marked with \verb|excluded_flag = 1| in the catalog, filtering out bursts with lower reliability or instrumental artifacts not suitable for statistical studies. Bursts with \verb|NaN| values for position or DM are also removed at the first step. We collapse multiple bursts from each repeating source into a single entry, retaining the first detected burst. This step guarantees that our sample consists of statistically independent bursts. 

Following a selection similar to that of \citet{Wang2025}, we require ${\rm DM_{MW}}<80\ {\rm pc\ cm^{-3}}$ and ${\rm DM_{exc}}>80\ {\rm pc\ cm^{-3}}$, where ${\rm DM_{exc}}={\rm DM_{obs}}-{\rm DM_{MW}}$. We estimate the Galactic ISM contribution with the NE2001 model \citep{Cordes2002}. The first cut limits the uncertainty in the Galactic DM estimate near the Galactic plane. The second cut accounts for a possible Milky Way halo contribution of $50-80\ {\rm pc\ cm^{-3}}$ \citep{Prochaska2019,Keating2020,Ravi2023}. We test different Galactic DM upper limits in Section~\ref{sec:galactic_tests}. These cuts leave 2656 FRBs before cluster exclusion and selection within the final $y$-map mask.

We also use the CIB-deprojected PR4 $y$-map of \citet{Chandran2026} for a paired map comparison. This reconstruction combines temperature and polarization data to reduce Galactic residuals and applies additional CIB constraints. We retain the original NILC map for the main analysis and describe the map comparison in Section~\ref{sec:cib_tests}.

\subsubsection{Additional cluster mask}
Since the tSZ signal is highly sensitive to electron temperature, it is predominantly driven by the hot intracluster medium (ICM) within virialized galaxy clusters. To reveal the faint signal from hotter WHIM gas in large-scale filaments and the diffuse IGM, we mask out known clusters. We mainly utilize two cluster catalogs: the second Planck catalogue of Sunyaev-Zeldovich sources \citep{Ade2016a} (PSZ2, containing 1653 targets) and the second release of the Meta-Catalogue of X-ray detected Clusters of galaxies \citep{Sadibekova2024} (MCXC-II, containing 2221 targets). For each cluster, we derive the physical radius $R_{500}$ based on the definition of $M_{500}$ and convert it to an angular radius $\theta_{500}$ using its redshift. The median angular radii for the PSZ2 and MCXC-II catalogs are $5.11'$ and $6.09'$, respectively.

We define exclusion regions based on multiples of $\theta_{500}$ (specifically adopting $1\theta_{500}$ as our baseline mask, and $3\theta_{500}$ for the subsequent robustness test) to mask both the FRB sample and the tSZ map. FRBs falling within these angular thresholds are directly discarded from the catalog. For the tSZ map, to prevent ringing artifacts in the Fourier domain caused by hard edges, we apply an additional apodization mask. We first generate a blank HEALPix map where the defined cluster exclusion regions are set to zero, and then apply a $C^1$ apodization using the \verb|NaMaster| package. To match the typical angular extent of the masked clusters, we adopt an apodization scale of $6'$. Because the \verb|NaMaster| apodization algorithm progressively smooths outwards from the mask boundaries into the valid sky regions, the inner exclusion zones remain strictly zero, guaranteeing the complete removal of ICM contamination. This newly generated apodized cluster mask is then multiplied by our base mask to produce the final combined mask for the cross-power spectrum estimation. We retain FRBs whose reported positions lie in pixels with nonzero weight in the final mask. The resulting samples contain 2035 and 1919 FRBs for the $1\theta_{500}$ and $3\theta_{500}$ masks, respectively. The final map we use is shown in Figure~\ref{fig:FRB_SZ_map}. For visual clarity, we show the version of $\theta_{\rm threshold}=3\theta_{500}$ only here.

\subsection{Power spectrum estimator}

We measure the angular cross-power spectrum $C_\ell$ between the FRB DM and tSZ fields as a function of the multipole moment $\ell$, corresponding approximately to an angular scale $\theta\sim180^\circ/\ell$. We use the pseudo-$C_\ell$ estimator in \verb|NaMaster| \citep{Alonso2019}. We provide the original $y$-map and the combined apodized mask to the spin-0 field \verb|NmtField| at $N_{\rm side}=2048$. The field applies the mask once. For the FRBs, we use the catalog-based field \verb|NmtFieldCatalog| with uniform source weights \citep{Wolz2025}. It evaluates the spherical harmonic transforms at the reported FRB positions without assigning the sources to map pixels. The field values are the excess DMs after subtracting the mean over the retained sample.

The incomplete sky coverage couples different multipoles. The coupled, or pseudo, cross-power spectrum $\tilde{C}_\ell$ is related to the underlying full-sky spectrum by $\langle\tilde{C}_\ell\rangle=\sum_{\ell'}M_{\ell\ell'}C_{\ell'}$,
where $M_{\ell\ell'}$ is the mode-coupling matrix associated with the map and catalog masks. We measure the coupled cross-power spectrum and compute this matrix with the \verb|NaMaster| workspace. We then invert the binned coupling matrix to obtain the decoupled bandpowers.

The initial measurement uses 20 logarithmically spaced bins over $40\leq\ell\leq6143$, with bin edges from 40 to 6144. The maximum included multipole is $\ell_{\rm max}=3N_{\rm side}-1=6143$. The low-$\ell$ modes are highly susceptible to large-scale bias and the variation of DM$_{\rm MW}$, as also discussed by \citet{Wang2025}. Consequently, our baseline analysis excludes the first bin, $40\leq\ell<51$, and uses the remaining 19 bins, with effective multipoles from 58 to 5459.5. The measured central values are calculated from the full retained sample.

We estimate the covariance using spatial jackknife resampling. A HEALPix grid with $N_{\rm side}=4$ divides the sky into 192 equal-area patches. To avoid poorly sampled regions, we use patches containing more than 10 retained FRBs, giving 53 and 52 jackknife regions for the $1\theta_{500}$ and $3\theta_{500}$ masks, respectively. In each realization, we remove the FRBs in one region and set the $y$-map mask to zero in the same region. We recalculate the mean DM of the remaining sources and recompute the mode-coupling matrix and decoupled spectrum. The covariance between bins $i$ and $j$ is
\begin{equation}
\Sigma_{ij}=\frac{N-1}{N}\sum_{r=1}^{N}
\left(C_{\ell_i}^{r}-\overline{C}_{\ell_i}\right)
\left(C_{\ell_j}^{r}-\overline{C}_{\ell_j}\right),
\end{equation}
where $C_{\ell_i}^{r}$ is the bandpower in bin $i$ after omitting region $r$, $N$ is the number of jackknife regions, and $\overline{C}_{\ell_i}$ is the mean over the $N$ realizations. The plotted uncertainties are $\sigma_i=\sqrt{\Sigma_{ii}}$. The retained bins have nonzero correlations, with maximum absolute off-diagonal correlation coefficients of 0.55 and 0.47 for the $1\theta_{500}$ and $3\theta_{500}$ masks, respectively. We retain the full covariance in the parameter inference.

\subsection{Theoretical framework}

The cross-correlation between DM and the $y$ parameter arises from the fact that both observables trace the same spatial distribution of free electrons. We model the intrinsic angular cross-power spectrum using the Limber approximation \citep{Limber1953,LoVerde2008}:
\begin{equation}
C_\ell^{yd}=f_{\rm WHIM}\int {\rm d}\chi\,
\frac{W_{\rm DM}(\chi)W_{\rm SZ}(\chi)}{\chi^2}
P_{ee}\left(k=\frac{\ell+1/2}{\chi},z(\chi)\right).
\label{eq:cross_theory}
\end{equation}
Here $\chi$ is the comoving distance and $k=(\ell+1/2)/\chi$ is the comoving wavenumber in ${\rm Mpc}^{-1}$. The electron overdensity power spectrum $P_{ee}$ has units of ${\rm Mpc}^3$, and $W_{\rm DM}$ and $W_{\rm SZ}$ are the radial weight functions. The fitted factor $f_{\rm WHIM}$ sets the amplitude of the hotter WHIM contribution, as defined below.

\subsubsection{The electron power spectrum}

We model the electron-density fluctuations with the gas profiles of \citet{Schneider2019}, implemented in \verb|BaryonForge| \citep{Anbajagane2024}. Assuming a uniform ionization fraction and composition, we use the gas overdensity spectrum as a template for $P_{ee}$. We calculate the mean comoving gas density, $\bar{\rho}_{\rm g}$, from the halo mass function and the gas masses obtained by integrating the same truncated profiles. This normalization gives the spectrum of $\delta_{\rm g}=\rho_{\rm g}/\bar{\rho}_{\rm g}-1$.

We calculate the one-halo and two-halo terms with the Core Cosmology Library \citep{Chisari2019}. The one-halo term contains the contribution from gas within the same halo. The two-halo term combines the gas profiles with the linear matter power spectrum and halo bias. We use the mass function of \citet{Tinker2008} and the halo bias of \citet{Tinker2010}. Halo masses are defined as $M_{200c}$ within $R_{200c}$, where the mean enclosed density is 200 times the critical density at that redshift. The mass integrals cover $10^9-10^{16}\ M_\odot$ and include the low-mass correction in the halo-model calculation. We suppress the one-halo term at $k\ll0.01\ {\rm Mpc}^{-1}$ to avoid a constant contribution on large scales.

Our main gas-profile parameters are $\theta_{\rm ej}=4$, $\theta_{\rm co}=0.1$, $M_c=10^{14}\ h^{-1}M_\odot$, $\mu_\beta=0.4$, $\gamma=2$, and $\delta=7$, where $h=H_0/(100\ {\rm km\ s^{-1}\ Mpc^{-1}})$. We retain gas beyond the halo radius and apply a truncation scale of $100R_{200c}$ for the numerical integration. This cutoff is separate from the angular cluster mask. We use the total-gas profiles as a spatial template, with the temperature and pressure response entering the SZ kernel below. Section~\ref{sec:gas_model_tests} tests the gas-profile parameters and the response to projected cluster holes.

\subsubsection{The tSZ weight function}

The tSZ signal traces electron pressure. We include a phenomenological response $B_{\rm pg}(z)$ to relate pressure fluctuations to the gas-density template. The density clustering is already described by $P_{ee}$, and $B_{\rm pg}$ accounts for the additional pressure response. The weight function is
\begin{equation}
W_{\rm SZ}(z)=W_{{\rm SZ},0}(1+z)^2B_{\rm pg}(z),\qquad
W_{{\rm SZ},0}=\frac{\sigma_{\rm T}k_{\rm B}T_e n_{e,0}}{m_e c^2}.
\label{eq:sz_kernel}
\end{equation}
We assume fully ionized hydrogen and helium with a helium mass fraction of 0.25. The present electron-density normalization is $n_{e,0}=(7/8)\Omega_b\rho_{{\rm crit},0}/m_p=2.20\times10^{-7}\ {\rm cm}^{-3}$, where $\rho_{{\rm crit},0}$ is the present critical density. The factor $(1+z)^2$ follows from the physical density scaling as $(1+z)^3$ and the proper path element ${\rm d}l={\rm d}\chi/(1+z)$.

We adopt a reference electron temperature $T_e=2.4\times10^6\ {\rm K}$ and a dimensionless pressure response $B_{\rm pg}(z)=6/(1+z)$, motivated by the gas-parameter model in \citet{Ibitoye2024}. The factor $B_{\rm pg}$ relates pressure fluctuations to the modeled gas-density fluctuations. We keep $T_e$ constant and fix both $T_e$ and $B_{\rm pg}$ during parameter inference. These values give $W_{{\rm SZ},0}=1.83\times10^{-10}\ {\rm Mpc}^{-1}$ and $W_{\rm SZ}(z)=1.10\times10^{-9}(1+z)\ {\rm Mpc}^{-1}$. Section~\ref{sec:discussion} discusses the physical interpretation of these assumptions.

We introduce $f_{\rm WHIM}=\Omega_{\rm WHIM}/\Omega_b$ as our primary fitting parameter, where $\Omega_{\rm WHIM}$ is the density parameter of the hotter WHIM probed by DM-tSZ. It represents the fraction of all cosmic baryons residing in this component under the adopted temperature and pressure model. Since $n_{e,0}$ is normalized to all cosmic baryons, $f_{\rm WHIM}$ linearly scales the cross-power spectrum in Equation~\ref{eq:cross_theory}. In contrast, $f_{\rm IGM}$ describes the total diffuse IGM fraction across thermal phases and enters the DM weight function below.

To compare with the measured cross-power spectrum, we include the SZ beam, pixel window, and FRB localization window:
\begin{equation}
C_\ell^{\rm th}=C_\ell^{yd}\times B_{\rm SZ}(\ell)
\times W^{\rm pix}(\ell)\times B_{\rm FRB}(\ell).
\label{eq:model_windows}
\end{equation}
Here $W^{\rm pix}(\ell)$ is the HEALPix pixel window at $N_{\rm side}=2048$, and $B_{\rm SZ}(\ell)=\exp[-\ell(\ell+1)\sigma_{\rm SZ}^2/2]$ is the Gaussian beam with $\sigma_{\rm SZ}=\theta_{\rm FWHM}/(2\sqrt{2\ln2})$ and $\theta_{\rm FWHM}=10'$. We describe $B_{\rm FRB}$ below.

\subsubsection{The FRB weight function}

The dispersion measure of an FRB is defined as the integrated column density of free electrons along the line of sight. For a cosmological FRB located at redshift $z$, the mean contribution from the IGM is given by the relation
\begin{equation}
\left\langle{\rm DM_{IGM}}\right\rangle= \frac{21 c  \Omega_{b} {H_0}^2 }{64 \pi H_0 G m_{p}} \times \int_{0}^{z} \frac{f_{\rm IGM} (1+z) {\rm d} z}{\left[\Omega_{m}(1+z)^{3} + 1 - \Omega_m \right]^{1 / 2}},
\end{equation}
where $m_p$ is the proton mass and $H_0$ is the Hubble constant. $\Omega_b$ and $\Omega_m$ are the baryon and matter density parameters, respectively. The parameter $f_{\rm IGM}$ represents the fraction of cosmic baryons residing in the diffuse IGM. We take the values of $\Omega_m, \Omega_b h^2$ and $H_0$ given by the Planck Collaboration \citep{Aghanim2020}, and assume $f_{\rm IGM} \approx 0.84$ \citep{Shull2012}. Both $f_{\rm WHIM}$ and $f_{\rm IGM}$ are normalized to all cosmic baryons. Within the diffuse-IGM interpretation of our model, the hotter WHIM probed by DM-tSZ is a subset of $f_{\rm IGM}$, which also includes cooler WHIM and photoionized gas \citep{Shull2012}. For consistent phase definitions, the ratio $f_{\rm WHIM}/f_{\rm IGM}$ describes the hotter component's share of the diffuse IGM, while $f_{\rm IGM}-f_{\rm WHIM}$ represents the fraction of all cosmic baryons in the remaining IGM. Since $f_{\rm IGM}$ is fixed rather than inferred jointly with $f_{\rm WHIM}$, this remainder is conditional on the adopted model and does not separately determine the cooler gas fractions.

For an ensemble of FRBs, the weight function at a given comoving distance $\chi$ depends not only on the local electron density but also on the probability that an FRB is located behind this redshift. Given a normalized FRB redshift probability density function, $p(z)$ (such that $\int_0^\infty p(z) dz = 1$), the weight function is derived by weighting the differential IGM contribution by the cumulative distribution of the FRB population. This expression is consistent with recent studies \citep{Wang2025,Leung2025,Sharma2025}.

Finite FRB localization precision suppresses the cross-power spectrum on small angular scales. We use the effective Gaussian window $B_{\rm FRB}(\ell)=\exp[-\ell^2/(2\ell_{\rm loc}^2)]$ in Equation~\ref{eq:model_windows} and fit $\ell_{\rm loc}$ jointly with $f_{\rm WHIM}$, following the approach of \citet{Wang2025}. Instrumental estimates for the first CHIME/FRB catalog give $315\lesssim\ell_{\rm loc}\lesssim1396$ \citep{RafieiRavandi2021}. The DM weight function is
\begin{equation}
W_{\rm DM}(\chi)=\frac{H(z)}{c}
\frac{{\rm d}\left\langle{\rm DM}_{\rm IGM}(z)\right\rangle}{{\rm d}z}
\int_z^{z_{\max}}p(z')\,{\rm d}z',\qquad z=z(\chi).
\end{equation}
This kernel contains $f_{\rm IGM}$ through the mean IGM DM. We examine residual cluster DM from localization errors and the dependence of the fitted amplitude on fixed localization scales in Section~\ref{sec:localization_tests}.

\subsubsection{The FRB redshift distribution}
\begin{figure}[t]
\centering
\includegraphics[width=0.7\textwidth]{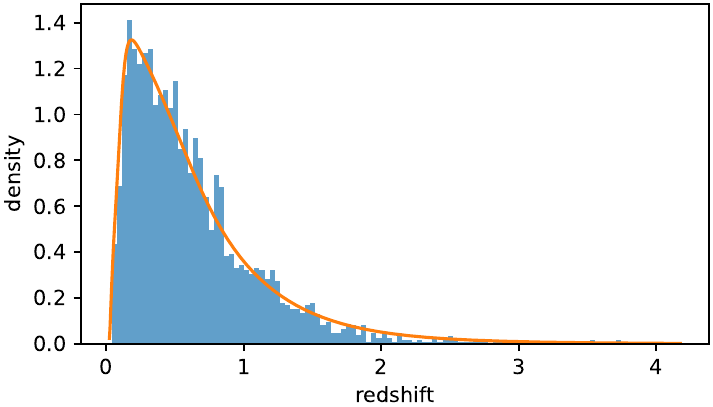}
\caption{\textbf{Redshift distribution of FRBs in CHIME/FRB Catalog 2.} Blue bars are a histogram of the median pseudo redshift of each FRB. The red curve is the interpolated redshift distribution given by smoothing the logarithmic KDE curve.}
\label{redshift}
\end{figure}
Conventional approaches to determining the FRB redshift distribution, $p(z)$, often rely on convolving an assumed FRB volumetric formation rate with a modeled instrumental selection function. To avoid the substantial model dependencies inherent in such parametrizations, we adopt a purely data-driven approach to construct $p(z)$ based on our observed sample. Following the methodology detailed in \citet{Gao2025}, we calculate the pseudo redshift posterior probability density for each non-localized FRB in our sample, and take the median of each probability density function (PDF) as the representative pseudo redshift of the source.

To derive a smooth and continuous probability density function from this discrete pseudo redshift catalog, while ensuring the physical boundary condition $p(z) \to 0$ as $z \to 0$, we perform a Kernel Density Estimation (KDE). To naturally enforce this origin constraint, the KDE is computed in the logarithmic domain ($\ln z$). The density is then properly normalized back to the linear $z$-space via the Jacobian transformation, $p(z) = p(\ln z) / z$. We adopt a smoothing bandwidth equal to twice the standard Scott's rule during the KDE to suppress small-scale fluctuations caused by finite sample size. Finally, to construct the continuous function required for the Limber integration, we resample the KDE curve using logarithmic spacing near the origin and linear spacing elsewhere, and apply a cubic spline interpolation. This interpolated curve is shown in Figure~\ref{redshift} and serves as the FRB redshift PDF in our cross-correlation analysis.

\begin{figure}[t]
\centering
\includegraphics[width=0.65\textwidth]{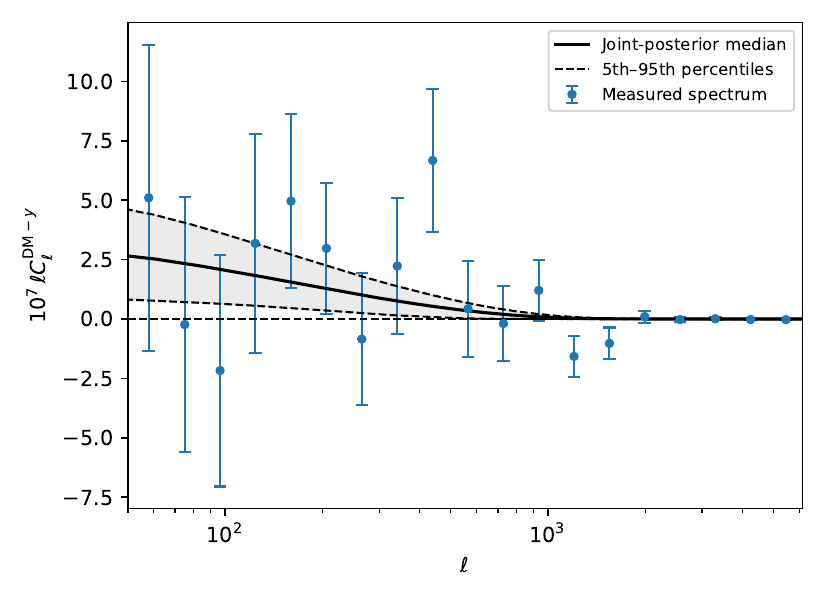} 
\caption{\textbf{Angular cross-power spectrum between the FRB dispersion measures and the tSZ Compton-$y$ parameter.} The blue points show the full-sample decoupled spectrum for the $1\theta_{500}$ mask, with jackknife uncertainties. The solid black line is the median model spectrum from the joint posterior samples of $f_{\rm WHIM}$ and $\ell_{\rm loc}$. The dashed lines and shaded region show the 5th-95th percentile interval, illustrating the posterior uncertainty envelopes.}
\label{power_spectrum}
\end{figure}

\subsection{Parameter inference and significance test}

To constrain the WHIM baryon fraction $f_{\rm WHIM}$ and the effective FRB localization scale $\ell_{\rm loc}$, we perform a Bayesian parameter inference by comparing the theoretical cross-power spectrum with the observational data. Assuming Gaussian uncertainties, we use the full jackknife covariance in the likelihood:
\begin{equation}
\ln\mathcal{L}=-\frac{1}{2}
\left({\boldsymbol C}^{\rm th}-{\boldsymbol C}^{\rm obs}\right)^{\rm T}
\boldsymbol{\Sigma}^{-1}
\left({\boldsymbol C}^{\rm th}-{\boldsymbol C}^{\rm obs}\right)+{\rm const}.
\end{equation}
Here ${\boldsymbol C}^{\rm obs}$ contains the full-sample measurements, and ${\boldsymbol C}^{\rm th}$ is evaluated at the effective multipoles, including the windows in Equation~\ref{eq:model_windows}. We hold the estimated covariance fixed during inference and evaluate the likelihood using a Cholesky decomposition of the correlation matrix.

We assign uniform priors $f_{\rm WHIM}\in\mathcal{U}(0,1)$ and $\ell_{\rm loc}\in\mathcal{U}(100,2000)$. We sample $\ell_{\rm loc}/1000$ to keep the parameter scales comparable. We use the MCMC ensemble sampler \verb|emcee| \citep{ForemanMackey2013}, with 128 walkers and 5000 steps. We estimate the integrated autocorrelation time $\tau$ for each parameter and discard the first $\lfloor\max(\tau)\rfloor+50$ steps. We report the marginalized posterior medians, with the 16th and 84th percentiles defining the central $68\%$ credible intervals.

We test the zero-cross-correlation hypothesis with the profile statistic $\Delta\chi^2=\chi^2_{\rm null}-\chi^2_{\rm best}$, using the same full covariance. We minimize $\chi^2$ over $0\leq f_{\rm WHIM}\leq1$ and a logarithmic grid of 512 localization scales between 100 and 2000. We calibrate this statistic with $10^5$ zero-mean Gaussian spectra drawn with the fixed jackknife covariance, repeating the same parameter search for each realization. The null tail probability is $p_{\rm null}=(n_{\geq\Delta\chi^2}+1)/(10^5+1)$, where $n_{\geq\Delta\chi^2}$ counts realizations with a statistic at least as large as observed. This calibration includes the localization-scale search and the amplitude boundary. We report the detection significance using $p_{\rm null}$, with $1-p_{\rm null}$ expressed as a confidence level.

\section{Results}
\label{sec:results}
\subsection{Measurements of $C_\ell$ and parameter inference}

The measured cross-power spectrum with $1\sigma$ error bars is shown in Figure~\ref{power_spectrum}. Note that to bring values at different multipole bins to the same magnitude, we plot $\ell C_\ell$ instead of $C_\ell$. As shown in the figure, we observe a broadly positive cross-correlation signal across the multipole range, indicating that FRBs and the tSZ effect trace the same underlying electron distribution.

\begin{figure}[t]
\centering
\includegraphics[width=0.6\textwidth]{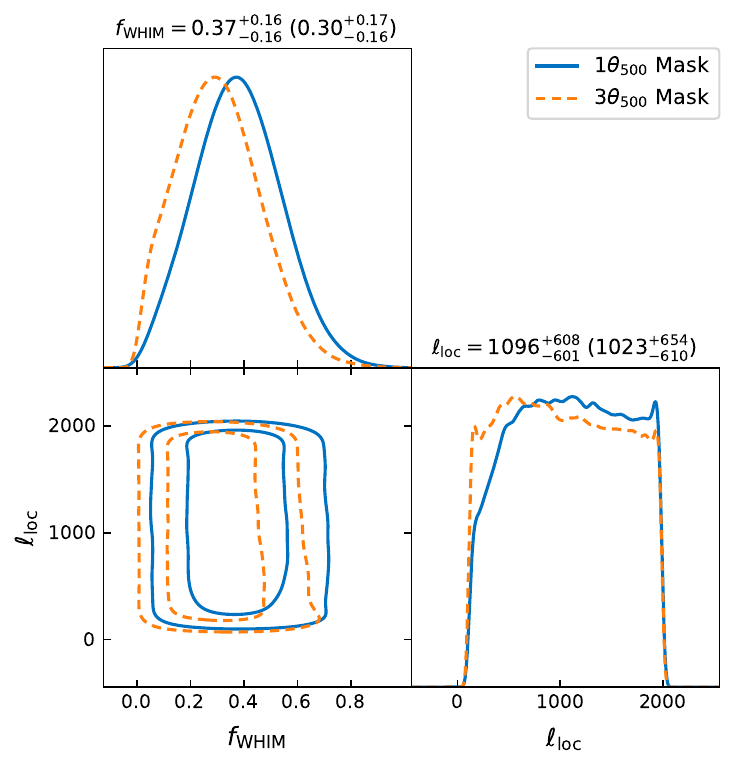}
\caption{\textbf{Posterior constraints on the hotter WHIM baryon fraction and FRB localization scale.} The plot shows the marginalized distributions and joint contours for the $1\theta_{500}$ mask (blue) and $3\theta_{500}$ mask (orange). For the $1\theta_{500}$ mask, the posterior medians and central $68\%$ credible intervals are $f_{\rm WHIM}=0.375^{+0.165}_{-0.161}$ and $\ell_{\rm loc}=1096^{+608}_{-601}$. The corresponding $3\theta_{500}$ results are $f_{\rm WHIM}=0.297^{+0.166}_{-0.156}$ and $\ell_{\rm loc}=1023^{+654}_{-610}$.}
\label{corner}
\end{figure}

We jointly fit $f_{\rm WHIM}$ and $\ell_{\rm loc}$ using the full jackknife covariance and the model in Equation~\ref{eq:model_windows}. We adopt the fiducial gas profiles, $T_e=2.4\times10^6\ {\rm K}$, and $B_{\rm pg}(z)=6/(1+z)$. Figure~\ref{corner} shows the posterior distributions. For the $1\theta_{500}$ mask, we obtain $f_{\rm WHIM}=0.375^{+0.165}_{-0.161}$ and $\ell_{\rm loc}=1096^{+608}_{-601}$, where the values are marginalized posterior medians and central $68\%$ credible intervals. In this model, the hotter WHIM contains roughly $37.5\%$ of cosmic baryons, providing a substantial contribution to the baryon census.

\subsection{Statistical significance and cluster-mask comparison}
For the $1\theta_{500}$ mask, the Gaussian null calibration gives $p_{\rm null}=0.0169$. We therefore detect a positive cross-correlation at $>98\%$ confidence. We also repeat the measurement with cluster exclusions expanded to $3\theta_{500}$ to examine the contribution from cluster outskirts. The larger mask reduces the retained sample and the available sky area. It gives $f_{\rm WHIM}=0.297^{+0.166}_{-0.156}$ and $\ell_{\rm loc}=1023^{+654}_{-610}$, with $p_{\rm null}=0.0631$. The hotter WHIM fractions inferred with the two masks agree within their uncertainties, with a lower central value for the larger mask. The difference is expected because a $3\theta_{500}$ mask naturally leads to less WHIM gas and thus a lower central value.

For comparison with Gaussian significance, the profile statistics are $\Delta\chi^2=5.264$ and 2.923 for the two masks. Their calibrated null probabilities correspond to $Z=2.12$ and 1.53, respectively, where $Z=\Phi^{-1}(1-p_{\rm null})$ and $\Phi$ is the standard normal cumulative distribution. These Gaussian-equivalent values summarize the same null test.

As a separate goodness-of-fit diagnostic, the minimum $\chi^2$ values are 21.38 and 19.06 for the two masks. With a nominal $19-2=17$ degrees of freedom, these give $p_{\rm gof}=0.210$ and 0.325, respectively. These approximate diagnostics use the standard $\chi^2$ distribution with the estimated jackknife covariance held fixed.

\section{Robustness and systematic tests}
\label{sec:systematics}

We examine the effects of gas-profile parameters and projected cluster holes in Section~\ref{sec:gas_model_tests}. Section~\ref{sec:localization_tests} tests residual cluster DM from localization errors and the sensitivity of the fits to fixed localization scales. We compare the NILC and CIB-deprojected CILC maps in Section~\ref{sec:cib_tests} and examine Galactic DM cuts and the NE2001 template in Section~\ref{sec:galactic_tests}. Sections~\ref{sec:host_tests} and \ref{sec:sky_selection_tests} test independent host-DM scatter and sky selection through right ascension randomization, respectively. Posterior medians and 68\% credible intervals are used for comparisons. The simulated cluster and host error bars describe scatter across realizations, while the observed error bars are from sky jackknife resampling.

\subsection{Gas-model sensitivity}
\label{sec:gas_model_tests}

We first test the dependence on the gas-profile parameters of \citet{Schneider2019} by taking $M_c=10^{13}$ and $10^{15}\ h^{-1}M_\odot$, or $\theta_{\rm ej}=3$ and 5, around the fiducial values. We vary one parameter at a time and keep the remaining gas-profile parameters, temperature, and pressure response fixed. Each fit uses the same 19 bins, full jackknife covariance, and priors as the main analysis, with $\ell_{\rm loc}$ free. Table~\ref{tab:gas_feedback} gives the marginalized constraints. The largest change in the posterior median is 0.039 for either mask, significantly smaller than the statistical uncertainty. The results are stable under these tested variants.

\begin{table}[htbp]
\centering
\caption{Sensitivity of the hotter WHIM fraction to the S19 gas-profile parameters. The first row is the fiducial model. Values are posterior medians with uncertainties from the 16th and 84th percentiles.}
\label{tab:gas_feedback}
\begin{tabular}{cccc}
\hline
$\theta_{\rm ej}$ & $M_c\ (h^{-1}M_\odot)$ & $f_{\rm WHIM}$ ($1\theta_{500}$) & $f_{\rm WHIM}$ ($3\theta_{500}$) \\
\hline
4 & $10^{14}$ & $0.375^{+0.165}_{-0.161}$ & $0.297^{+0.166}_{-0.156}$ \\
4 & $10^{15}$ & $0.406^{+0.182}_{-0.176}$ & $0.324^{+0.185}_{-0.170}$ \\
4 & $10^{13}$ & $0.336^{+0.145}_{-0.141}$ & $0.258^{+0.148}_{-0.136}$ \\
3 & $10^{14}$ & $0.340^{+0.146}_{-0.143}$ & $0.259^{+0.149}_{-0.134}$ \\
5 & $10^{14}$ & $0.404^{+0.179}_{-0.175}$ & $0.321^{+0.185}_{-0.170}$ \\
\hline
\end{tabular}
\end{table}

We next test the response of the theoretical cross-power spectrum to the cluster mask. While calculating the theoretical spectrum using \verb|BaryonForge|, we model this response by removing central projected gas from halos that are selected to match the PSZ2 and MCXC-II catalogs \citep{Ade2016a,Sadibekova2024}. We convert the catalog $M_{500c}$ estimates to $M_{200c}$ with the same dark-matter profile as in the S19 model. In each mass-redshift cell, we set the selection probability to $S(M,z)=N_{\rm catalog}/N_{\rm halo}$, where $N_{\rm halo}$ is calculated from the fiducial mass function over the same sky area.

For selected halos, we subtract the central projected profile within $1\theta_{500}$ or $3\theta_{500}$ on the DM side. On the $y$ side, we also include a radial $6'$ $C^1$ transition to remain consistent with the observed map. We retain the outer gas, the original mean gas density, and the low-mass correction. The one-halo term averages the products for the same selected or unselected halo,
\begin{equation}
\langle u_{\rm D}u_y\rangle_S
=(1-S)u_{\rm D}^{0}u_y^{0}
+S u_{\rm D}^{\rm hole}u_y^{\rm hole},
\end{equation}
where $u_{\rm D}$ and $u_y$ are the gas-density profiles entering the DM and $y$ integrals.

\begin{figure}[htbp]
\centering
\includegraphics[width=0.5\textwidth]{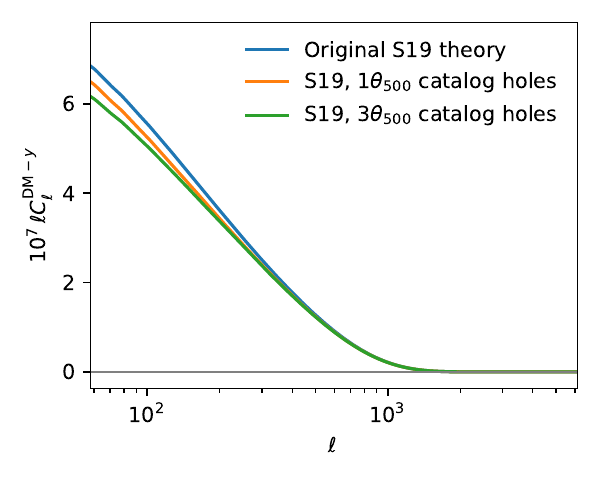}
\caption{\textbf{Response of the theoretical cross-power spectrum to projected cluster holes.} The curves show the fiducial S19 template (blue) and the templates with catalog-matched $1\theta_{500}$ (orange) and $3\theta_{500}$ (green) holes. For comparison, all curves use $f_{\rm WHIM}=1$ and $\ell_{\rm loc}=1000$, with the same SZ beam and pixel window.}
\label{fig:theory_mask_response}
\end{figure}

Figure~\ref{fig:theory_mask_response} compares these templates with the fiducial spectrum. Replacing the fiducial template in the same parameter inference gives $f_{\rm WHIM}=0.398^{+0.175}_{-0.170}$ and $0.322^{+0.185}_{-0.168}$ for the $1\theta_{500}$ and $3\theta_{500}$ masks, respectively. The median shifts are 0.0233 and 0.0255, about 14\% and 15\% of the respective upper $68\%$ uncertainty widths of the main results. This comparison supports a small response for the modeled catalog population under the adopted profiles.

\subsection{FRB localization and residual cluster DM}
\label{sec:localization_tests}

FRB localization errors can affect the measurement in two ways. They smooth the cross-power spectrum, which we describe with the effective Gaussian window in Equation~\ref{eq:model_windows}. They can also leave an FRB in the retained sample even when its true line of sight passes through a cluster exclusion region. A freely fitted $\ell_{\rm loc}$ could partly absorb scale-dependent suppression associated with the cluster mask. We examine these effects with a residual cluster-DM simulation and fits at fixed localization scales.

\begin{figure}[htbp]
\centering
\includegraphics[width=\textwidth]{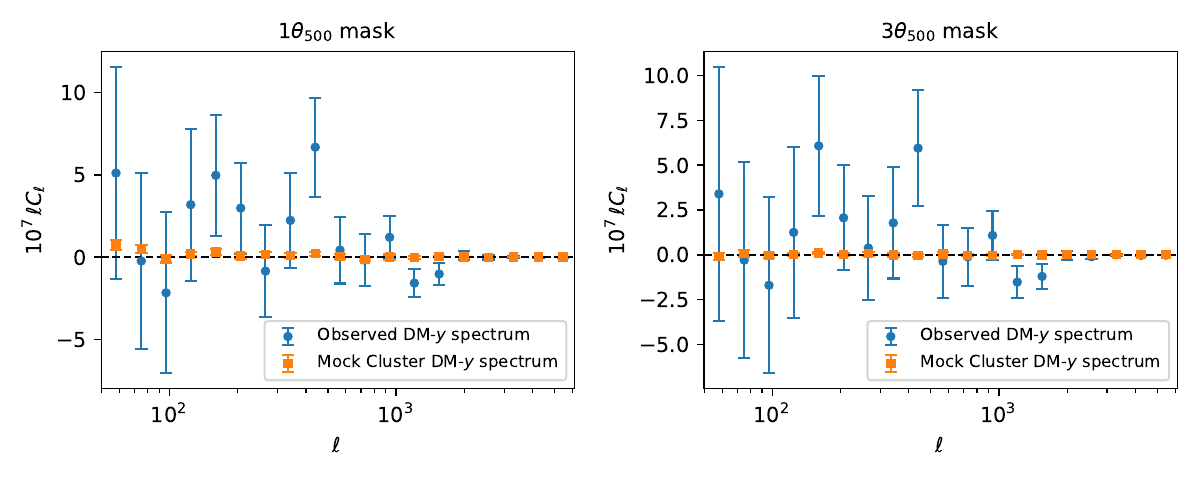}
\caption{\textbf{Residual cluster DM from FRB localization errors.} Blue points show the current observed DM-$y$ spectra, and orange points show the means of 32 simulated cluster-DM spectra. Orange error bars are the standard deviations across the localization realizations. The left and right panels use $1\theta_{500}$ and $3\theta_{500}$ masks, respectively.}
\label{fig:cluster_dm_localization}
\end{figure}

We first estimate the residual cluster-DM signal while accounting for FRB localization errors. We generate 32 realizations of the true FRB positions by drawing independent Gaussian RA and declination errors, treating the reported errors from CHIME/FRB Catalog 2 as $1\sigma$. Sources shared by the two mask samples use the same position draw. We calculate the cluster DM along each drawn line of sight using the PSZ2 and MCXC-II catalogs. For the 2742 distinct clusters with valid physical parameters, we adopt a spherical $\beta$-model,
\begin{equation}
n_e(r)=\frac{n_0}{1+(r/r_c)^2},\qquad r<3R_{500},
\end{equation}
and set the density to zero at larger radii. We choose $\beta=2/3$ and $r_c=0.1R_{500}$, and normalize $n_0$ to $M_{\rm gas}(<R_{500})=0.12M_{500}$. These parameters define a rough diagnostic cluster model. We further assume that all FRBs lie behind all intersected clusters. To remain consistent with the fiducial measurement, the simulated cluster DM is assigned to the original reported FRB position before measuring its cross-power spectrum. Thus the true position determines the electron column, while the reported position determines the measured angular correlation. Figure~\ref{fig:cluster_dm_localization} shows the mean simulated spectra and their standard deviations across the 32 realizations. Both the absolute mean signal and the uncertainty of the cluster spectrum are significantly smaller than the observed spectrum and its uncertainty. This comparison shows the residual cluster-DM signal is relatively small in our result.

We next test sensitivity to the effective localization scale, which is weakly constrained by the main fit. We repeat the fits with $\ell_{\rm loc}$ fixed at 400, 1000, and 2000, using the same prior on $f_{\rm WHIM}$. Table~\ref{tab:fixed_localization} compares these results with the baseline fits in which $\ell_{\rm loc}$ is free. The maximum changes in the posterior median of $f_{\rm WHIM}$ are 0.0065 and 0.0181 for the two masks, respectively, much smaller than their statistical intervals. The inferred amplitude therefore depends weakly on the selected effective smoothing scale over this range. Together with the residual-DM comparison, we show that the FRB localization error has little influence on our parameter inference. The effective localization scale does not absorb mask-related suppression since its value has little impact on the $f_{\rm WHIM}$ fit.

\begin{table}[htbp]
\centering
\caption{$f_{\rm WHIM}$ constraints at fixed FRB localization scales. Values are marginalized posterior medians with central $68\%$ credible intervals.}
\label{tab:fixed_localization}
\begin{tabular}{lcc}
\hline
$\ell_{\rm loc}$ & $f_{\rm WHIM}$ ($1\theta_{500}$ mask) & $f_{\rm WHIM}$ ($3\theta_{500}$ mask) \\
\hline
Free (baseline) & $0.375^{+0.165}_{-0.161}$ & $0.297^{+0.166}_{-0.156}$ \\
400 & $0.375^{+0.166}_{-0.163}$ & $0.305^{+0.171}_{-0.159}$ \\
1000 & $0.377^{+0.161}_{-0.158}$ & $0.291^{+0.160}_{-0.152}$ \\
2000 & $0.368^{+0.160}_{-0.157}$ & $0.279^{+0.158}_{-0.147}$ \\
\hline
\end{tabular}
\end{table}

\subsection{CIB deprojection and map comparison}
\label{sec:cib_tests}

\begin{figure}[htbp]
\centering
\includegraphics[width=\textwidth]{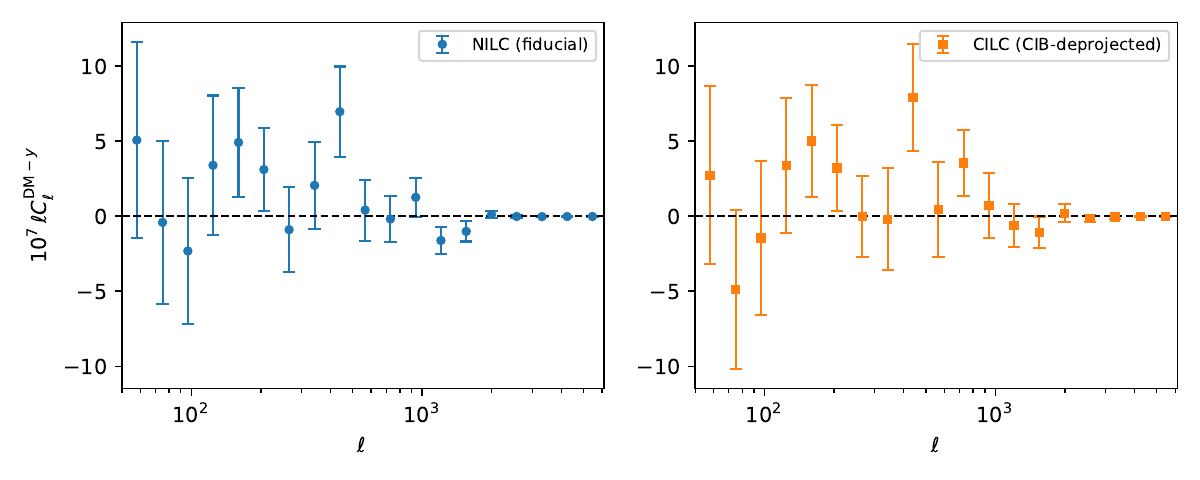}
\caption{\textbf{NILC and CILC map comparison.} The DM-$y$ cross-power spectra are measured with the PR4 NILC map (left) and the CIB-deprojected CILC map (right). Both use the $1\theta_{500}$ cluster exclusion. Central points are the full-sample estimates, and error bars are from 53 paired sky-jackknife regions.}
\label{fig:nilc_cilc}
\end{figure}

We test how the $y$-map reconstruction affects the measurement by comparing the fiducial PR4 NILC map with the CIB-deprojected map of \citet{Chandran2026}. The latter combines temperature and polarization data to reduce Galactic residuals and deprojects the zeroth-order CIB moment and its first-order temperature moment. We use the $1\theta_{500}$ cluster exclusion, and 2035 FRBs, with 53 paired sky-jackknife regions. Figure~\ref{fig:nilc_cilc} shows the two cross-power spectra. Using the same theoretical template and full-covariance likelihood within this comparison, we obtain $f_{\rm WHIM}=0.395^{+0.163}_{-0.160}$ for NILC and $0.509^{+0.185}_{-0.186}$ for CILC. Admittedly, there is a deviation between the median values, but the marginalized intervals overlap. The CILC map uses a different reconstruction, and it also changes the treatment of Galactic foregrounds other than CIB, so the difference cannot be attributed to CIB deprojection alone. CIB intensity, measured in ${\rm MJy\ sr^{-1}}$, cannot be directly compared with dimensionless $y$ to estimate a contamination fraction without a map-specific leakage coefficient. As a result, we retain the original NILC map for the main analysis.

\subsection{Galactic foregrounds}
\label{sec:galactic_tests}
We first test sensitivity to the Galactic DM selection by varying the upper limit on ${\rm DM_{MW}}$ from 60 to 100~${\rm pc\ cm^{-3}}$, while retaining ${\rm DM_{exc}}>80~{\rm pc\ cm^{-3}}$. We repeat the spectrum, covariance, and parameter estimation for each selection with the same map, cluster mask, and theoretical model. Table~\ref{tab:dm_mw} lists the results for the two cluster masks. The 60 and 100~${\rm pc\ cm^{-3}}$ selections agree with the fiducial 80~${\rm pc\ cm^{-3}}$ constraints within their uncertainties. A stricter 40~${\rm pc\ cm^{-3}}$ cut retains only 997 and 921 FRBs for the $1\theta_{500}$ and $3\theta_{500}$ masks, respectively. These samples provide only 25 and 24 jackknife regions for the 19-bin covariance. Fits with this stricter cut are sensitive to the cluster mask and can approach the prior boundary, so we do not use them to establish parameter stability.

\begin{table}
\centering
\caption{Galactic DM upper-cut tests. $f_{\rm WHIM}$ results are posterior medians with 16th-84th percentile intervals. The 80~${\rm pc\ cm^{-3}}$ row is the fiducial result.}
\label{tab:dm_mw}
\begin{tabular}{ccc}
\hline
${\rm DM_{MW}}$ upper cut & $f_{\rm WHIM}$ ($1\theta_{500}$) & $f_{\rm WHIM}$ ($3\theta_{500}$) \\
$({\rm pc\ cm^{-3}})$ & & \\
\hline
60 & $0.423^{+0.171}_{-0.170}$ & $0.365^{+0.176}_{-0.170}$ \\
80 & $0.375^{+0.165}_{-0.161}$ & $0.297^{+0.166}_{-0.156}$ \\
100 & $0.441^{+0.169}_{-0.166}$ & $0.382^{+0.181}_{-0.174}$ \\
\hline
\end{tabular}
\end{table}

\begin{figure}[htbp]
\centering
\includegraphics[width=\textwidth]{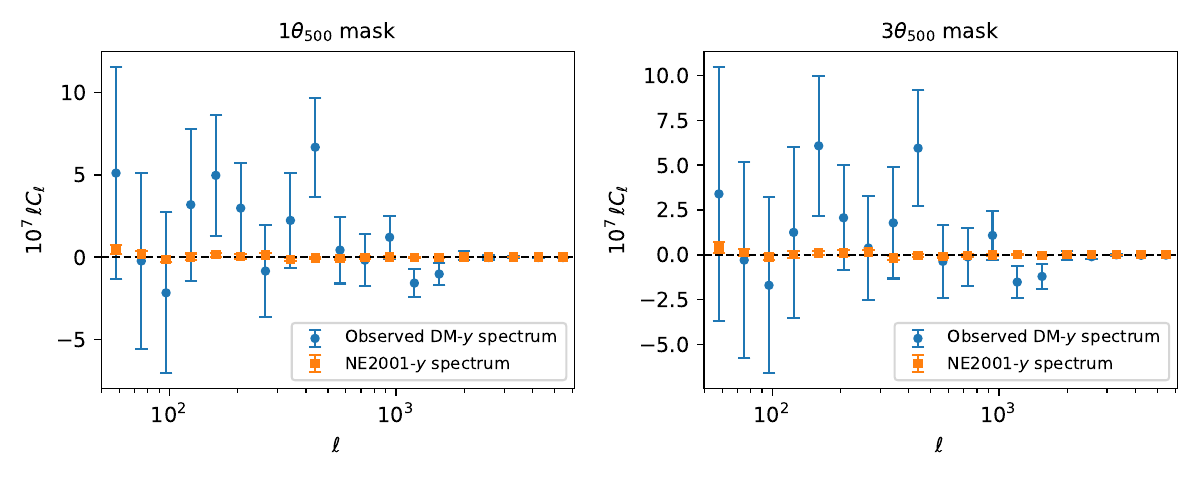}
\caption{\textbf{Galactic DM template test.} Observed DM-$y$ cross-power spectra and NE2001-$y$ spectra for the $1\theta_{500}$ (left) and $3\theta_{500}$ (right) cluster masks. Central values are full-sample means, and error bars are sky-jackknife uncertainties.}
\label{fig:ne2001}
\end{figure}

In a second Galactic foreground test, we cross-correlate the NE2001 Galactic DM predictions \citep{Cordes2002} at the retained FRB positions with the masked $y$-map. Figure~\ref{fig:ne2001} compares these spectra with the observed DM-$y$ spectra. The largest absolute Galactic contribution is 0.072 and 0.057 times the observed single-bin uncertainty for the two masks. The absolute mean signal and the uncertainty of the Galactic foreground spectrum are significantly smaller than the observed spectrum and its uncertainty.

\subsection{Host-galaxy DM scatter}
\label{sec:host_tests}
Though host galaxies themselves do not have a contribution to FRB cosmological DM, it is possible that host galaxies are correlated with large-scale filaments, and thus correlated with the SZ map. However, with current CHIME FRB data, we could not locate the host galaxies, and the correlation between host galaxies and filaments cannot be quantified with a convincing model.

As a result, we could only estimate the accidental cross-correlation from host-DM scatter. We keep the 2035 FRB positions in the fiducial $1\theta_{500}$ sample and independently draw a host contribution from a log-normal distribution. We use three fixed pairs $(e^\mu,\sigma_{\rm host})$: $(36.55,1.27)$ from the non-repeating-FRB model at $z=0.1$ of \citet{Zhang2020}, $(85.11,0.944)$ from \citet{Shin2023}, and $(163.23,0.91)$ from the TNG100-1 PSFR-VNum model of \citet{Mo2023}. Here $e^\mu$ is in ${\rm pc\ cm^{-3}}$, and $\sigma_{\rm host}$ is dimensionless. We use these as fixed-amplitude scatter scenarios, without assigning source redshifts or applying a $1/(1+z)$ conversion. They are not calibrated observer-frame distributions for this sample.

\begin{figure}[htbp]
\centering
\includegraphics[width=\textwidth]{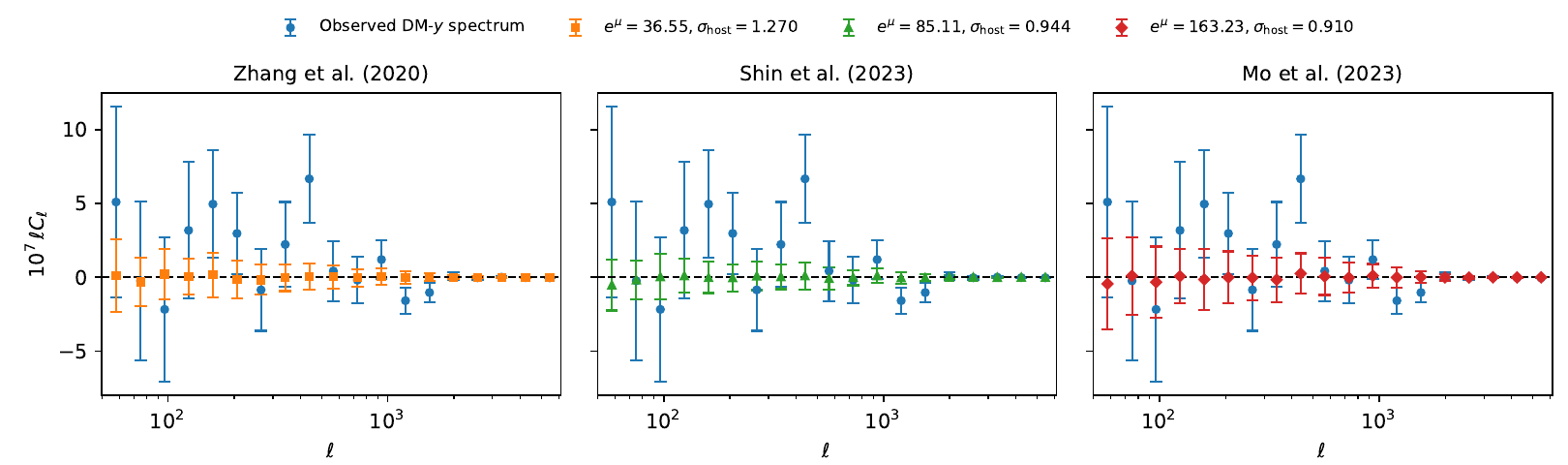}
\caption{\textbf{Host-galaxy DM scatter tests.} The panels use three log-normal parameter sets. Each panel compares the observed DM-$y$ spectrum with the mean of 100 mock host-DM spectra using the $1\theta_{500}$ cluster mask. Mock error bars show the standard deviation across realizations. Observed error bars are the fiducial sky-jackknife uncertainties.}
\label{fig:host_dm}
\end{figure}

For each parameter pair, we calculate 100 realizations. Figure~\ref{fig:host_dm} shows the mean spectra and their standard deviations between realizations. The absolute mean contribution remains much lower than the observed spectrum, while the host uncertainties, unlike those in tests in previous subsections, are not negligible due to the long-tailed nature of the $\rm DM_{host}$ distribution.

\subsection{Sky-selection randomization}
\label{sec:sky_selection_tests}

\begin{figure}[htbp]
\centering
\includegraphics[width=0.5\textwidth]{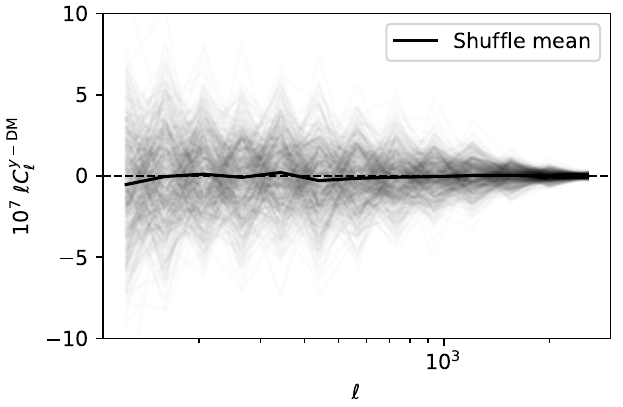}
\caption{\textbf{Right ascension randomization test.} The semi-transparent curves show the cross-power spectra of 512 randomized catalogs, and the black curve shows their mean. Declination and DM are held fixed while right ascension is randomized within the mask.}
\label{shuffle}
\end{figure}

Finally, we use right ascension randomization to test for false correlations caused by the non-uniform CHIME sky selection. We keep each FRB's declination and DM fixed and draw a new right ascension, accepting only positions within the final mask. This preserves the observed declination distribution while removing the physical spatial correspondence with the $y$ field. Figure~\ref{shuffle} shows the spectra of 512 randomized catalogs and their mean. The mean randomized spectrum is consistent with zero in all bins, and its absolute amplitude is substantially smaller than that of the measured cross-power spectrum. This null test provides no evidence that the observed signal is produced by the non-uniform CHIME sky coverage or detection sensitivity.

\begin{figure}[h]
\centering
\includegraphics[width=0.7\textwidth]{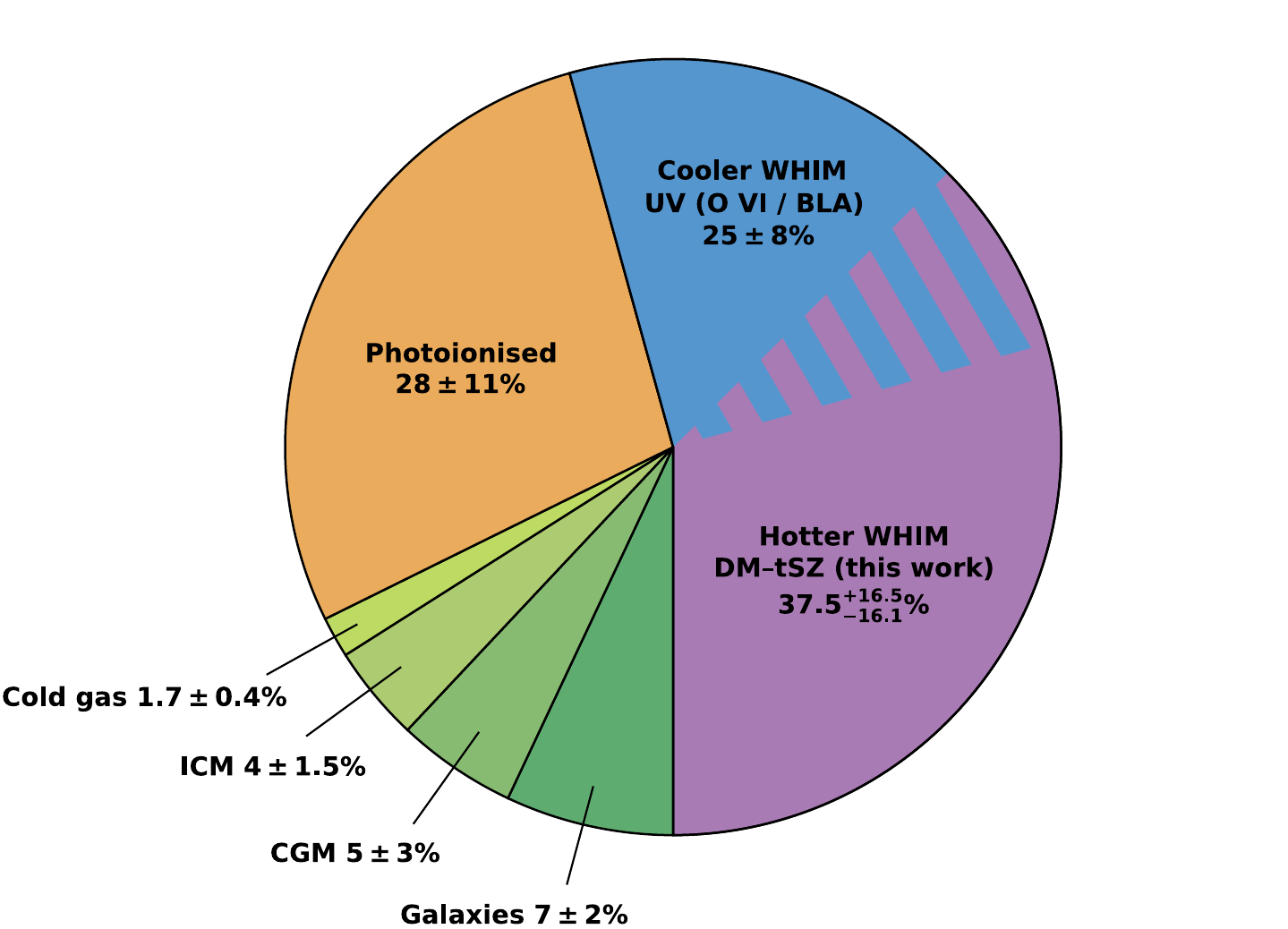}
\caption{\textbf{The statistically closed cosmic baryon census.} Labels give the fraction of cosmic baryons and its uncertainty for each component. Literature estimates follow \citet{Shull2012}. Our DM-tSZ result for the hotter WHIM is $37.5^{+16.5}_{-16.1}\%$, giving the posterior median and central $68\%$ credible interval for the $1\theta_{500}$ mask under the adopted temperature and pressure response. Purple denotes the hotter WHIM probed by DM-tSZ, while blue denotes WHIM traced by UV absorption \citep{Shull2012}. The interlocking region visually illustrates possible overlap between the two tracers, rather than a sharp temperature boundary.}
\label{baryon}
\end{figure}

\section{Discussion and Conclusion}
\label{sec:discussion}

In this work, hotter WHIM describes the gas preferentially weighted by the temperature-sensitive tSZ signal, without imposing a sharp temperature boundary. At fixed gas profiles and redshift evolution, the cross-power spectrum amplitude scales as $f_{\rm WHIM}T_eB_{{\rm pg},0}$, where $T_e=2.4\times10^6\ {\rm K}, B_{{\rm pg},0}=6$ in our model. The inferred baryon fraction therefore scales inversely with the adopted temperature and pressure normalization, which are not separately determined by this measurement. For example, \citet{Ibitoye2024} constrained the product of gas bias, electron density, and temperature. Adopting a gas bias of 6, they inferred $T_e\simeq2.4\times10^6\ {\rm K}$. The temperature and bias thus form a paired choice and motivate our reference temperature and pressure normalization. Note that our measurement can include contributions from gas in undetected halos, cluster outskirts beyond the exclusion radii, and voids, but does not separate these contributions by environment. For other parameters and models, the tests in Section~\ref{sec:systematics} support stability of the inferred amplitude under the tested S19 parameters, localization scales, catalog-matched profile holes, and Galactic DM cuts of $60-100\ {\rm pc\ cm^{-3}}$. The map comparison and independent host simulations provide additional foreground diagnostics.

By combining FRB electron columns with thermal-pressure information, the DM-tSZ cross-correlation probes the hotter diffuse baryons that are difficult to detect directly \citep{Munoz2018}. After masking cataloged galaxy clusters, we detect a positive cross-correlation at $>98\%$ confidence ($p_{\rm null}=0.0169$). With the adopted temperature and pressure normalization, the inferred fraction is $f_{\rm WHIM}=0.375^{+0.165}_{-0.161}$. This result supports a substantial baryon reservoir in the hotter WHIM and complements observations of cooler gas through UV absorption \citep{Shull2012}. The hotter WHIM traced here may also include gas redistributed beyond halo boundaries by feedback \citep{Veenema2026}. Figure~\ref{baryon} places the hotter WHIM constraint in the current baryon census. The literature estimates include galaxies ($7\pm2\%$), cold gas ($1.7\pm0.4\%$), the CGM ($5\pm3\%$), the ICM ($4\pm1.5\%$), and photoionized Ly$\alpha$ gas ($28\pm11\%$) \citep{Shull2012}. The underlying observations include inventories of stars and intracluster gas \citep{Fukugita2004}, 21-cm surveys of neutral hydrogen \citep{Zwaan2003}, and absorption-line measurements of circumgalactic gas \citep{Tumlinson2011}.

Our DM-tSZ fraction of $37.5^{+16.5}_{-16.1}\%$ probes a hotter gas component, complementing the $25\pm8\%$ traced by O~VI and broad Ly$\alpha$ absorbers \citep{Shull2012}. These tracers can include some of the same gas, though this overlapping gas component may not contribute much to our DM-tSZ signal due to lower temperature and pressure. As a result, the two estimates cannot simply be added as independent components. The interlocking colors illustrate this possible overlap rather than a measured division between thermal phases. These probes extend the baryon census across different thermal conditions and help account for the missing baryons in the local Universe. Together, the baryon budget is statistically closed under our adopted density and pressure model.

\section*{acknowledgments}
This work was supported by the National Natural Science Foundation of China (grant Nos. 12625304, 12494575, and 12393812).

\vspace{5mm}
\bibliography{reference}{}
\bibliographystyle{aasjournal}

\end{document}